\documentclass[aps, prb, 10pt, superscriptaddress, twocolumn]{revtex4-1}

\usepackage{amsmath,amssymb}
\usepackage{graphicx}

\newcommand{\la}{\label}
\newcommand{\bbm}{\begin{multline}}
\newcommand{\eem}{\end{multline}}
\newcommand{\be}{\begin{equation}}
\newcommand{\ee}{\end{equation}}
\newcommand{\bea}{\begin{eqnarray}}
\newcommand{\eea}{\end{eqnarray}}
\newcommand{\p}{\partial}

\newcommand{\comment}[1]{}
\usepackage{color}

\allowdisplaybreaks[1]

\begin{document}

\title{Electromagnetic and gravitational responses of two-dimensional non-interacting electrons in background magnetic field}

\author{Alexander G.~Abanov}
\affiliation{Department of Physics and Astronomy, Stony Brook University,  Stony Brook, NY 11794, USA}
\affiliation{Simons Center for Geometry and Physics,
Stony Brook University,  Stony Brook, NY 11794, USA}
%\affiliation{Department of Physics and Astronomy and Simons Center for Geometry and Physics,
%Stony Brook University,  Stony Brook, NY 11794, USA}

\author{Andrey~Gromov}
\affiliation{Department of Physics and Astronomy, Stony Brook University,  Stony Brook, NY 11794, USA}

\date{\today}

%%%%%%%%%%%%%%%%%%%%%%%%%%%%%%%%%%%%%%%%%%%%%%%%%%%%%%%%%%%%%%%%%%%%%%%%%%%%%%%%%%%%%%%%%%%%%%%
\begin{abstract}
We compute electromagnetic, gravitational and mixed linear response functions of two-dimensional free fermions in external quantizing magnetic field at an integer filling factor. The results are presented in the form of the effective action and as an expansion of currents and stresses in wave-vectors and frequencies of the probing electromagnetic and metric fields. We identify the terms in linear response functions coming from geometric Chern-Simons, Wen-Zee, and gravitational Chern-Simons terms in effective action.  We derive the expressions for Hall conductivity, Hall viscosity and find the current and charge density responses to the spatial curvature as well as stresses caused by inhomogeneous electromagnetic fields. 
\end{abstract}

%\pacs{03.65.Ud}{Entanglement and quantum nonlocality} -- CORRECT
%\pacs{05.30.Fk}{Fermion systems and electron gas}
%\pacs{05.40.-a}{Fluctuation phenomena, random processes, noise, and Brownian motion} -- CORRECT

%%%%%%%%%%%%%%%%%%%%%%%%%%%%%%%%%%%%%%%%%%%%%%%%%%%%%%%%%%%%%%%%%%%%%%%%%%%%%%%%%%%%%%%%%%%%%%%

\maketitle

%%%%%%%%%%%%%%%%%%%%%%%%%%%%%%%%%%%%%%%%%%%%%%%%%%%%%%%%%%%%%%%%%%%%%%%%%%%%%%%%%%%%%%%%%%%%%%%
%%%%%%%%%%%%%%%%%%%%%%%%%%%%%%%%%%%%%%%%%%%%%%%%%%%%%%%%%%%%%%%%%%%%%%%%%%%%%%%%%%%%%%%%%%%%%%%
\section{Introduction}
\label{sec:intro}
%%%%%%%%%%%%%%%%%%%%%%%%%%%%%%

Recent interest to the Hall viscosity in the theory of Fractional Quantum Hall effect (FQHE) and the interest to the interplay of defects and mechanical stresses with electromagnetic properties of materials motivates studies of gravitational, electromagnetic and mixed responses in condensed matter physics. Gravitational field in condensed matter systems can be understood either as a way to represent deformational strains present in the material under consideration or as a technical tool allowing to extract correlation functions involving stress tensor components.

It is always important to have a simple model system for which such responses can be calculated exactly. For the quantum Hall effect one can consider two-dimensional electron gas in a constant magnetic field (2DEGM) as such a model. When the density of fermions is commensurate with magnetic field the integer number of Landau levels is filled and one expects local and computable response to weak external fields. This model is as important starting point of analysis for quantum Hall systems as a free electron gas for the theory of metals. However, while some electromagnetic responses for 2DEGM can be found in literature we were not able to find the complete results for mixed and gravitational linear responses. The goal of this paper is to compute these responses providing the analogue of Lindhard\cite{fetter-walecka-book} function, both e/m and gravitational, for 2DEGM. We compute the effective action encoding linear responses in the presence of external inhomogeneous, time-dependent, slowly changing electro-magnetic and gravitational fields.

We compare and find an agreement of the obtained responses with known e/m responses \cite{1989-ChenWilczekWittenHalperin, Salam1990CS-SC, lykken1990anyonic,wiegmann1990parity} and with known results for Hall viscosity at integer fillings \cite{2009-Read-HallViscosity, bradlyn-read-2012kubo}. In addition we find the stress, charge and current densities induced by spatial curvature. Another point of comparison is given by phenomenological hydrodynamic models for FQHE \cite{1986-GirvinMacDonaldPlatzman,1989-Read,1989-ZHK,1990-Stone,wiegmann2013hydrodynamics,abanov2013FQHE}  and Ward identities following from the exact local Galilean symmetry (also known as non-relativistic diffeomorphism) of the model\cite{2011-HoyosSon, 2006-SonWingate} . 

The paper is organized as follows. In Section \ref{sec-eff-action} we describe the model as a non-relativistic quantum field theory and present our results in terms of the effective action.  In Section \ref{sec-em-resp} we extract the electromagnetic responses from the effective action. We demonstrate some peculiar physical effects such as charge accumulation/depletion in the presence of conic singularity in metric,  non-dissipative current perpendicular to a gradient of curvature and report higher gradient corrections to Hall conductivity. Our main results are presented in the Section \ref{sec-grav-resp} where we discuss the gravitational responses. We present higher gradient and dynamic corrections to Hall viscosity.  We leave the in-depth discussion of dynamic responses and their relation to the local Galilean invariance for a separate publication.

%%%%%%%%%%%%%%%%%%%%%%%%%%%%%%%%%%%
\section{Effective action}
\la{sec-eff-action}
%%%%%%%%%%%%%%%%%%%%%%%%%%%%%%%%%%%

%%%%%%%%%%%%%%%%%%%%%%%%%
\subsection{The model}
%%%%%%%%%%%%%%%%%%%%%%%%%

Our starting point is the system of two-dimensional non-interacting and non-relativistic fermions interacting with an external gauge $A_{\mu}$ and spatial metric $g_{ij}$ fields. We assume that there is no curvature of space-time $g_{00}=g_{0i}=0$, but that the spatial metric can depend on time. The action has a form
\begin{multline}
 \la{action}
	S =\int d^2x dt \sqrt{g}\left[ \frac{i}{2}\hbar \psi^{\dag}\p_0\psi 
	-\frac{i}{2}\hbar( \p_0\psi^{\dag})\psi + \right.   
 \\
	+ \left. eA_0\psi^\dag\psi-\frac{\hbar^2}{2m}g^{ij}(D_i\psi)^{\dag}D_j\psi
	+\frac{g_s B}{4m}\psi^\dag\psi\right] \,.
\end{multline}
Here we assumed that the fermions are spin polarized and treat $\psi$ field as a complex grassman scalar. We have also added Zeeman term with the g-factor $g_s$. For the case of electrons in vacuum $g_s=2$, but it is convenient to keep it arbitrary for potential condensed matter applications. The covariant derivative $D_i = \p_i - i\frac{e}{\hbar c}(\bar{A}_i+A_i)$ and includes both vector potential of the constant background magnetic field $B_{0}=\p_{1}\bar{A}_{2}-\p_{2}\bar{A}_{1}$ and a weak perturbation. In the curved background magnetic field is defined as $B=\frac{1}{\sqrt{g}}\left(\p_1\bar{A}_2 - \p_2\bar{A}_1 + \p_1A_2 - \p_2A_1 \right)$, so it transforms as a (pseudo)scalar under coordinate transformations. We separate it into constant part and perturbation as $B=B_0 + b$. In this work we use the expression linear in fields $b=B-B_{0}\approx \p_1A_2 - \p_2A_1-\frac{1}{2}\delta g_{ii}$. Here $\delta g_{ij}$ is a deviation from the flat background $g_{ij} = \delta_{ij} + \delta g_{ij}$.

We omit the chemical potential term in (\ref{action}) for brevity, but assume throughout the paper that the lowest $N$ Landau levels are completely filled in the ground state.
We use conventional notations for metric fields so that $g_{ij}$ and $g^{ij}$ are reciprocal matrices and an invariant spatial volume is given by $\sqrt{g}\,d^2x$ with $g=\det(g_{ij})$. 

Finding linear responses of the system (\ref{action}) with respect to varying gauge and metric fields amounts to the computation of the effective action of the theory in quadratic (aka RPA) approximation. The effective action $S_{eff}$ is defined as a path integral over the fermionic fields
\be
 \la{seff}
	e^{\frac{i}{\hbar}S_{eff}[A_\mu,g_{ij}]} 
	\equiv \int  D(g^{\frac{1}{4}}\psi) D(g^{\frac{1}{4}}\psi^\dag) e^{\frac{i}{\hbar}S} \,.
\ee
The notation\cite{fujikawa1980anomaly,hawking1977zeta} $D(g^{\frac{1}{4}}\psi)$ serves as a reminder that the path integral is taken over the space of functions $\psi(x),\psi^\dagger(x)$ equipped with the invariant scalar product given by
\be
	(\psi,\phi) \equiv \int d^2x\, \sqrt{g}\psi^\dag\phi \,.
\ee 

%%%%%%%%%%%%%%%%%%%%%%%%%
\subsection{The effective action}
%%%%%%%%%%%%%%%%%%%%%%%%%

The effective action defined in (\ref{seff}) can be computed as a regular expansion in background fields $A_\mu(x,t)$ and $g_{ij}(x,t)$ and their gradients. In the following 
we expand the effective action to quadratic order in fields. It is convenient to separate it as
\be
	S_{eff} = S_{eff}^{(1)}+S_{eff}^{(geom)}+S_{eff}^{(2)} \,.
 \la{seff-split}
\ee
The first contribution is given by
\be
 \label{seff-1}
	S_{eff}^{(1)} = \int d^2x dt \sqrt{g}\left[\epsilon_0 + \rho_0 A_0 +s_0 \omega_0\right] \,,
\ee
where $\omega_0$ is the time component of the spin connection (see explanations after (\ref{serf-top})) and $\epsilon_0$, $\rho_0$, and $s_0$ are the energy density, density, and orbital spin density in the ground state are given respectively by
\be
	\rho_{0} = \frac{N}{2\pi l^{2}}\,, 
	\quad
	\epsilon_{0} = \rho_{0}\,\hbar \omega_{c}\frac{2N-g_{s}}{4}\,, 
	\quad
	s_0=\rho_0\hbar \frac{N}{2} \,.
\ee
Here and throughout the paper we use conventional notations for magnetic length and cyclotron frequency given in term of the constant part of the background magnetic field $B_{0}$ as
\be
	l^{2} =\frac{\hbar c}{e B_{0}}\,, 
	\qquad
	\omega_{c} = \frac{eB_{0}}{mc}\,.
\ee
We notice here that although (\ref{seff-1}) includes all terms linear in $A_\mu,\, g_{ij}$ they also contain some of quadratic terms. Indeed the expansion of the metric factor in terms of the deviations from the flat background is
\be
	\sqrt{g} = 1 + \frac{1}{2}\delta g_{ii} -\frac{1}{8}\left[(\delta g_{11}-\delta g_{22})^{2}+4\delta g_{12} \delta g_{21}\right] +\ldots
\ee
and (\ref{seff-1}) should be re-expanded in truncated to terms up to the second order in fields.

The second term in (\ref{seff-split}) contains the topological and geometrical contributions to the effective action (with $\hbar=c=e=1$) 
\bea
	S_{eff}^{(geom)} &=&  \frac{N}{4\pi}\int
	\left(
	AdA
	+ N A d \omega
	+\frac{2N^2-1}{6} \omega d\omega
	\right)\,,
 \la{serf-top}
\eea
where we used the ``form notation'' $\int AdA \equiv \int d^2x dt \epsilon^{\mu\nu\lambda}A_\mu\partial_\nu A_\lambda$ etc. The coefficients of the three terms in (\ref{serf-top}) give, respectively the Hall conductivity $\sigma_{H} = \frac{N}{2\pi}$, the average orbital spin per particle $\bar{s}=\frac{N}{2}$ (corresponding to the Wen-Zee shift ${\cal S}=N$), and the gravitational Chern-Simons (gCS) coefficient $\frac{N(2N^{2}-1)}{24\pi}$.

The following comment is in order. The action (\ref{action}) is written in terms of the gauge potential $A_\mu$ and metric $g_{ij}$. It does not require spin connection $\omega_\mu$ as it is already covariant due to the fact that $\psi$ is a scalar field. Thus, the $S^{(geom)}_{eff}$ should also depend solely on the vector potential and metric. It is, however, instructive to write $S^{(geom)}_{eff}$ in terms of $A_\mu$ and $\omega_\mu$ as in (\ref{serf-top}). We used the fact that the gCS term could be written exactly solely in terms of metric up to the boundary terms. \cite{Jackiw-2003-Reduced_gCS,Stone-Gravitational}.
With the accuracy used in this paper going back to metric in (\ref{serf-top}) amounts to 
$\omega_i \leftrightarrow -\frac{1}{2}\epsilon^{jk}\partial_j \delta g_{ik}$ and $\omega_0 \leftrightarrow \frac{1}{2}\epsilon^{jk}\delta g_{ij}\delta \dot{g}_{ik}$. 

It is illuminating to present (\ref{serf-top}) as an explicit sum over Landau levels
\bea 
 	S_{eff}^{(geom)} =  \sum_{n=1}^{N} \int
	\Big[
	\frac{1}{4\pi}(A+\bar{s}_n\omega)d(A+\bar{s}_n\omega)
 \nonumber \\
	- \frac{c}{48\pi} \omega d\omega
	\Big] \,, \qquad c=1\,,
 \la{s-ill}
\eea
where $\bar{s}_n = \frac{2n-1}{2}$ is the orbital spin per particle on the $n$-th Landau level and the last term is an anomalous gCS contribution the same for all Landau levels. It is equal to the non-relativistic limit of the gCS \cite{Stone-Gravitational} and corresponds to the $c=1$ CFT on the boundary. Its presence shows that the spin connection does not simply combine with vector potential in the effective action (cf. Refs.~\onlinecite{WenZeeShiftpaper},\onlinecite{1993-frohlich}). 

The physical meaning of Chern-Simons and Wen-Zee terms have been extensively discussed in literature. The gravitational Chern-Simons term is less known and is usually related to the transverse heat transport via Luttinger's argument. \cite{2000-ReadGreen,Stone-Gravitational,nishioka2012chiral,2010-RyuMooreLudwig,wang2011topological}  Additionally, it describes the induction of the angular momentum by curvature and the fractional anyonic statistics of conical singularities (disclinations). Here we consider only the relation of the gCS term to the Hall viscosity (see Sec.~\ref{subsec-stress}).

Finally, the last term in (\ref{seff-split}) gives the remaining second order terms
\bea
	S_{eff}^{(2)} &=& \int d^{2}x dt\, \mathcal{L}^{(2)} \,,
 \la{seff-nontop} \\
 	\mathcal{L}^{(2)} &=& \frac{1}{2}\left(A_\mu \Pi^{\mu\nu}A_\nu 
	+ A_\mu\Theta^{\mu}_{ij}\delta g^{ij}
	+\delta g^{ij}\Lambda_{ijkl}\delta g^{kl}\right) \,,
 \nonumber
\eea
where differential operators $\Pi,\Theta,\Lambda$ encode electro-magnetic, mixed, and gravitational responses, respectively. These operators can be computed exactly as infinite series in time and spatial derivatives or as series in frequency and wavevectors in Fourier representation.  We will present the details of the computation elsewhere and give here only the results obtained in the lowest orders in gradients 
\bea
	\frac{4\pi}{N} \mathcal{L}^{(2)} 
	&=&  ml^2E_{i}^2-\frac{N}{m}b^2 - \frac{3N}{2} l^2 b (\p_iE_{i}) 
	+ \frac{2N^2-1}{4m}bR  \,,
 \nonumber \\
 	&+&   \frac{2N^2-1}{6} l^2 R (\p_iE_{i}) +\frac{N(N^2-1)}{8m}R^2 +\ldots \,.
  \la{seff-2} 
\eea
Here $R$ is the scalar curvature given by $R=\p_i\p_j \delta g_{ij} - \Delta \delta g_{ii}$ after linearization.
While the first three terms of the expansion (\ref{seff-2}) can be found in literature\cite{1989-ChenWilczekWittenHalperin} the other terms present a result of this work.

The effective action presented above is, probably, the most compact way to summarize linear responses. However, we find it convenient to have direct formulas for observables such as charges, currents and stresses in a dynamic and inhomogeneous background. We present the explicit expressions and their physical meaning for linear responses in next sections.
For the illustration purposes and to lighten up the equations in the following we consider only the lowest Landau level filled, i.e. $N=1$.

%%%%%%%%%%%%%%%%%%%%%%%%%
\section{Electromagnetic responses}
 \la{sec-em-resp}
%%%%%%%%%%%%%%%%%%%%%%%%%

The expectation values of electric charge density and current are given by variational derivatives of the action (\ref{seff}) with respect to scalar and vector gauge potentials, respectively.  We have
\bea
	\rho(x)  &\equiv & \frac{1}{\sqrt{g}} \frac{\delta S_{eff}}{\delta A_0(x)}
	=\langle \psi^\dag\psi \rangle \,,
 \nonumber \\
	 J^i(x)  &\equiv &  \frac{1}{\sqrt{g}}\frac{\delta S_{eff}}{\delta A_i(x)}
	= \left\langle \frac{1}{2m i}g^{ij}
	\left[ \psi^\dag D_j\psi - (D_j\psi)^\dag\psi\right] \right\rangle  \,.
 \nonumber
\eea
Here the averaging is performed over the ground state in inhomogeneous background with integer number of Landau levels filled.

%%%%%%%%%%%%%%%%%%%%%%%%%
\subsection{Density}
%%%%%%%%%%%%%%%%%%%%%%%%%
We start with static response to an inhomogeneous scalar potential $A_0$. In the curved background the density has to be understood as number of particles per invariant volume element 
\begin{multline} \la{rho}
 	\rho-\rho_0
	=\frac{1}{2\pi}\left(1+\frac{3-g_s}{4}l^2\Delta\right) b\\
 	+\frac{1}{8\pi}\left(1+\frac{1}{3}l^2\Delta\right)R
 	+\frac{ml^2}{2\pi}\left(1+\frac{3}{8}l^2\Delta\right)(\partial_i E_i)\,,
\end{multline}
where $\Delta$ is the Laplacian. Notice, that (\ref{rho}) gives the response of the density to the curvature $\p\rho/\p R = \frac{1}{8\pi}\left(1-\frac{1}{3}|kl|^2+\ldots\right)$ (cf., Ref.~\onlinecite{wiegmann-abanov-2013}). Integrating (\ref{rho}) over a closed manifold we obtain the shift in the total charge due to a topology of the manifold
\be
	Q= \int d^2x \sqrt{g}\rho = \int d^2x\sqrt{g} \left(\frac{B}{2\pi} + \frac{R}{8\pi}\right)
	= N_\phi + \frac{1}{2}\chi \,,
 \la{Qshift}
\ee
where $N_\phi$ is the total magnetic flux and $\chi$ is the Euler characteristics of the manifold.\cite{WenZeeShiftpaper}
The correction to density due to curvature gradients in (\ref{rho}) 
is in agreement with  Refs.~\onlinecite{wiegmann-abanov-2013, abanov2013FQHE}. Extending (\ref{Qshift}) to the case of an isolated conic singularity with the deficit angle $\theta$ we find
\be
	\delta Q = \int d^2 x\sqrt{g}\left(\rho-\rho_0\right) = \frac{1}{8\pi}\int d^2x \sqrt{g} R
	=\frac{1}{4\pi}\theta \,.
 \la{coneQ}
\ee
The points of higher positive curvature suck particles in and increase local density. Although the derivation presented here cannot be rigorously applied to the case of conic singularity where curvature $R=2\theta\delta(x)$ is highly singular, the integral formula (\ref{coneQ}) is exact and can be checked by direct computation of the density on a surface of the cone.

%%%%%%%%%%%%%%%%%%%%%%%%%%
%\subsubsection{Dynamics}
%%%%%%%%%%%%%%%%%%%%%%%%%%

Detailed discussion of the dynamic response functions requires an analysis of the local Galilean invariance and is beyond the scope of this paper. In the following we illustrate some structures arising as the time dependence is introduced. 

In the flat background (no metric perturbations) and for $N=1$, $g_{s}=0$ we have
\bea
	\frac{\rho(\omega)}{\rho_0}&=&\frac{1}{1-\omega^2}
	\Big(1+l^2b+ml^4\p_iE_i 
 \nonumber \\
 	&-& \frac{3}{2}l^2\Delta\frac{2l^2b+ml^4\p_iE_i}{4-\omega^2}
	+\ldots\Big) \,,
 \nonumber
\eea
where $\omega$ is measured in units of $\omega_{c}$.
 The overall pole at $\omega=1$ is expected even in the presence of interactions as a consequence of the Kohn's theorem. The poles at $\omega=n$, $n=2,3,\ldots$ corresponding to transitions between different Landau levels occur in the next terms of gradient expansion. It is instructive to compare coefficients of $b$ and $ml^2\p_iE_i$ in the leading and subleading terms. In the leading order the coefficients are the same and the combination is Galilean invariant. In the next order the relative coefficient is different and the local Galilean transformation of the term will generate gravitational terms. 

Expanding in frequency and including the gravitational perturbations we have the leading term (first order in time derivative)
\be
 	\rho(x,t) = \rho(x,0)  + \frac{3}{16\pi ml^2}\epsilon_{ij}\p_i\p_k\dot{g}_{ik} 
\ee
with $\rho(x,0)$ given by (\ref{rho}).
It turns out that this expression is covariant with respect to the local Galilean transformation, i.e. local time-dependent coordinate transformations \cite{2006-SonWingate}. This happens due to intricate relations between the coefficients in the gradient expansion (Ward identities). 

%%%%%%%%%%%%%%%%%%%%%%%%%
\subsection{Electric current}
%%%%%%%%%%%%%%%%%%%%%%%%%
Response to the vector potential is given by
\be
	\langle J_i\rangle = \epsilon^{ij}\left(\sigma_H E_{j} 
	+ \frac{2-g_s}{4\pi m}\partial_j\left(b  +\frac{R}{8}\right)\right) \,,
 \la{jresp}
\ee
where the wavevector dependent Hall conductivity is given by
\be\la{sigma}
	\sigma_H(k) = \frac{1}{2\pi}\left(1-\frac{3-g_s}{4}|kl|^2+\frac{22 - 9g_s}{96}|kl|^4\right)\,.
\ee
The correction of the order of $k^2$ is in full agreement with general results for Galilean invariant systems \cite{bradlyn-read-2012kubo,2011-HoyosSon}. The $k^4$ term calculated here shows that $g_s$ term in the action (\ref{action}) gives a higher order contribution to the wavevector dependent Hall conductivity.

The second term in (\ref{jresp}) is a one of the results of this work. It shows that in low orders of gradient expansion the gradient of magnetic field and curvature affect current similarly to the electric field.
We also point out that in agreement with\cite{son2013newton} zero mass limit is regular for $g_s=2$.

%%%%%%%%%%%%%%%%%%%%%%%%%
\section{Gravitational responses}
 \la{sec-grav-resp}
%%%%%%%%%%%%%%%%%%%%%%%%%

Similarly to the electromagnetic responses the expectation value of stress tensor is given as 
\bea
	 T_{ij}  \equiv  -\frac{2}{\sqrt{g}}\frac{\delta S_{eff}}{\delta g^{ij}(x)} 
	 &=& \frac{1}{2m} \Big\langle
	\left(D_i\psi)^\dag D_j\psi + (D_j\psi)^\dag D_i\psi \right) 
	\Big\rangle
 \nonumber \\
 	&-& \frac{1}{4m}g_{ij} (\Delta_g+g_s B)
	\left\langle  \psi^\dag\psi  \right\rangle
 \la{Tij} \,.
\eea
Here $\Delta_g$ is the Laplace-Beltrami operator defined as $\Delta_g\rho=\frac{1}{\sqrt{g}}\partial_i ( g^{ij}\sqrt{g}\partial_j\rho)$. 
\footnote{We remark here that the term $-\frac{1}{4m}g_{ij} \Delta_g\rho$ in (\ref{Tij}) comes from the path integral measure (\ref{seff}) while the rest of (\ref{Tij}) can be obtained in conventional way from the variation of (\ref{action}) over the metric.}

%%%%%%%%%%%%%%%%%%%%%%%%%%%%
\subsection{Stress tensor}
\la{subsec-stress}
%%%%%%%%%%%%%%%%%%%%%%%%%%%%

Using (\ref{Tij}) we find the stress tensor in the leading order in gradients
\bea
	T_{ij} &=& \frac{1}{8\pi} \left(\p_i E_j + \p_j E_i\right) 
  \la{Tijres} \\
	&+& \delta_{ij} \left(\epsilon_0 -\frac{4-g_s}{8\pi}\p_k E_k
	+ \frac{2-g_s}{8\pi ml^2}\left(b + \frac{R}{8}\right) \right) \,.
 \nonumber
\eea

As before, stress tensor has regular limit $m\rightarrow 0$ limit for $g_s=2$.

Trace of the metric tensor $\delta g_{ii}$ couples to the Hamiltonian and thus describes the correction to the energy density due to the inhomogeneous gravitational potential. This is true only in the leading order in gradient expansion since in the next to the leading order there is an additional contribution coming from $\Delta \rho$ as can be seen from (\ref{Tij}). Keeping only the lower gradients we obtain the correction to the energy density
\be
 	\epsilon - \epsilon_0 = -\frac{4-g_s}{8\pi}\p_iE_i 
	+ \frac{2-g_s}{8\pi ml^2}\left(b + \frac{R}{8}\right)\,.
\ee
In the case of an isolated conic singularity we get a contribution to the total energy $\frac{\delta E}{E_0} =\frac{\theta}{8\pi} $ per singularity.\footnote{$E_0=\epsilon_0/\rho_0$ is the energy per particle in the unperturbed state.}

Time-dependent part of the stress tensor is related to another quantity of great interest: the Hall viscosity. 
We are looking for the parity odd terms in the stress tensor containing no more than two spatial derivatives. 
\bea
	T^{odd}_{ij} &=& \frac{1}{2}\eta_H (\epsilon_{ik}\dot{g}_{kj} 
	+ \epsilon_{jk}\dot{g}_{ki}) 
 \la{Todd}\\
	&+&  \frac{1}{2}\eta^{(2)}_H l^2\left[\epsilon_{il}\partial_l\partial_j 
	+ \epsilon_{jl}\partial_l\partial_i\right]\dot{g}_{kk}
 \nonumber 
\eea
where $\eta_H(\omega, k)$ can be considered as a frequency and wavevector dependent Hall viscosity (here $N=1$ and we measure $\omega$ in units of $\omega_c$) 
\be
	\frac{\eta_H(\omega,k)}{\eta_H^{(0)}} 
	= \frac{4}{4-\omega^2} 
	+ |kl|^2\left(\frac{1}{1-\omega^2}-\frac{6}{4-\omega^2}+\frac{6}{9-\omega^2}\right)\,.
 \nonumber
\ee
Here the conventional Hall viscosity
\be
	\eta_H(\omega=0,k=0)\equiv \eta_H^{(0)} = \frac{1}{2}\rho_0\bar{s} \,.
\ee
Notice, that at zero wavevector $\eta_H(\omega)/\eta_H^{(0)} = 4/(4-\omega^2)$ in full agreement with Ref.~\onlinecite{bradlyn-read-2012kubo}. 
For the coefficient in front of the second tensor (second line of Eq.~\ref{Todd}) we have
\bea
	\eta_H^{(2)} &=&  \frac{1}{8}\rho_0\left(\frac{2}{1-\omega^2} -\frac{4}{4-\omega^2}\right)\,.
\eea

In the static limit and for general $N$ we rewrite the expression for the Hall viscosity as a sum over Landau levels
\be\la{eta2}
	\eta_H(k) = \frac{1}{2\pi l^2}\sum_{n=1}^N \left(\frac{\bar{s}_{n}}{2} 
	+\frac{1}{4} \left[\bar{s}_{n}^2-\frac{c}{12}\right]|kl|^2\right) \,.
\ee
One has to recall that $c=1$ and that the orbital spin per particle at the $n$-th Landau level $\bar{s}_{n} = \frac{2n-1}{2}$.
We remark here that the gCS term gives a long wave $k^2$ correction to the Hall viscosity in a fashion similar to how the Wen-Zee term produces the long wave correction to the Hall conductivity. \cite{2011-HoyosSon} In fact,  in the non-interacting model considered in this work the $k^2$ correction to the Hall viscosity (\ref{eta2}) comes solely from the gCS term.

We note, that the gCS term also corrects the value of the Hall viscosity in the presence of constant background curvature $R_0$.  Indeed, the gCS term gives a contribution $\sqrt{g}\frac{1}{2}R_0\omega_0$ to the effective action, which results in $\delta \eta_H = \frac{N(2N^2-1)}{96\pi}R_0$. Then the total value of the Hall viscosity is given by
\be \la{etaR}
	\eta_H = \frac{1}{2\pi l^2} \sum_{n=1}^N \left(\frac{\bar{s}_n}{2} 
	+ \frac{1}{8} \left[\bar{s}_n^2-\frac{c}{12}\right]R_0l^2 \right)\,.
\ee
The second term gives the correction due to the curvature of the background and should be compared to (\ref{eta2}). Equation (\ref{etaR}) is in (somewhat surprising) correspondence with Ref.~\onlinecite{hughes2012torsion}, where the same (for $N=1$) curvature-induced shift of the relativistic version of the Hall viscosity was observed.

%%%%%%%%%%%%%%%%%%%%%%%%%%%%%%%%%
\section{Conclusions}
%%%%%%%%%%%%%%%%%%%%%%%%%%%%%%%%%
We have considered non-interacting two-dimensional fermions in the background electromagnetic and metric fields (\ref{action}). We have computed the effective action in the second order in deviations from the background of flat metric and constant magnetic field for an integer filling factor. The results are presented both in terms of the effective action and as linear response formulas for density, current and stress. The effective action features geometric Chern-Simons, Wen-Zee, and gravitational Chern-Simons terms. 
The higher gradient corrections to Hall conductivity and Hall viscosity have been computed. 
 
The model considered is known to be of fundamental importance for the understanding of the quantum Hall effect and topological phases of matter.  Our results provide a good starting point for refined variational and hydrodynamic approaches to QHE and elucidate phenomenological and symmetry based relations between linear responses in Galilean invariant systems found recently. 
A possible generalization of this work is to include torsion into the background in order to analyze responses to dislocations.\footnote{The role of torsion was discussed before \cite{hughes2011torsional,hughes2012torsion} where it was shown that the Hall viscosity appears as a response to torsion in relativistic theory. }

When the paper has already been completed we learned about the work\cite{CLW-tobe} where some of the results presented here have been extended to FQHE.

We thank B. Bradlyn, M. Goldstein, T. Hughes, K. Jensen, G. Monteiro, P. van Nieuwenhuizen, and P. Wiegmann  for helpful discussions. The work of A.G.A. was supported by the NSF under grant no. DMR-1206790.

\bibliographystyle{my-refs}

%\bibliographystyle{plain}
%\bibliographystyle{unsrt}
%\bibliographystyle{iopart-num}
%\bibliographystyle{abbrv}
%\bibliographystyle{apsrev}
%\bibliographystyle{amsplain}
%\bibliographystyle{ieeetr}
%\bibliographystyle{apsrev4-1}
%\bibliographystyle{aipnum4-1.bst}

%\bibliographystyle{unsrtnat}

%\begin{thebibliography}{10}

%%%%%%%%%%%%%%%%%%%%
\bibliography{abanov-bibliography}
%%%%%%%%%%%%%%%%%%%%
%%%%%%%%%%%%%%%%%%%%%%%%%%%%%%%%%%%%%%%
%%%%%%%%%%%%%%%%%%%%%%%%%%%%%%%%%%%%%%%
%%%%%%%%%%%%%%%%%%%%%%%%%%%%%%%%%%%%%%%

\end{document}